\documentclass[twocolumn,showpacs,preprintnumbers,amsmath,amssymb]{revtex4}

\usepackage{epsfig}

\begin{document}


\title{The equivalence of two approaches to the design of entanglement distillation protocols.}

\author{Erik Hostens}
\email{erik.hostens@esat.kuleuven.ac.be}
\affiliation{Katholieke Universiteit Leuven, ESAT-SCD, Belgium}
\author{Jeroen Dehaene}
\affiliation{Katholieke Universiteit Leuven, ESAT-SCD, Belgium}
\author{Bart De Moor}
\affiliation{Katholieke Universiteit Leuven, ESAT-SCD, Belgium}
\date{\today}

\begin{abstract}
  We show the equivalence of two approaches to the design of
entanglement distillation protocols. The first approach is based on
local unitary operations that yield permutations of tensor products
of Bell states. The second approach is based on stabilizer codes.
\end{abstract}

\pacs{03.67.Mn, 03.67.Pp, 03.67.Hk}

\maketitle

\newcommand{\phiplus}{|\Phi^{+}\rangle}

\newcommand{\psiplus}{|\Psi^{+}\rangle}

\newcommand{\phimin}{|\Phi^{-}\rangle}

\newcommand{\psimin}{|\Psi^{-}\rangle}

\newcommand{\plusphi}{\langle\Phi^{+}|}

\newcommand{\pluspsi}{\langle\Psi^{+}|}

\newcommand{\minphi}{\langle\Phi^{-}|}

\newcommand{\minpsi}{\langle\Psi^{-}|}
\newcommand{\psitilde}{\tilde{\Psi}}

\section{Introduction}
We describe a link between two approaches to the design of
entanglement distillation protocols. The first approach, which we call
{\em permutation based}, has it origins in Refs.~\cite{BBP:96,B:96},
where protocols are presented based on local operations that permute
the 16 possible tensor products of 2 Bell states. This approach
was generalized in Ref.~\cite{DVD:03} through a description in binary
arithmetic of all locally realizable permutations of all tensor
products of $n$ Bell states. Several other protocols based on similar ideas
as in Refs.~\cite{BBP:96,B:96} have been proposed. We only mention
Refs.~\cite{DEJ:96,MaS:99,Met:01,VV:04}. The second
approach, which we call {\em code based}, derives distillation
protocols from stabilizer codes as in Refs.~\cite{M:03,AmG:03}.

In the permutation based approach the protocol consists of unitary operations applied by Alice and
Bob to their side of a number of qubit pairs (in some of the references above also higher
dimensional systems are considered, but in this paper we will restrict ourselves to qubits),
followed by a number of measurements on some of the pairs (single qubit measurements). The
measurement outcomes are compared to decide if the protocol has succeeded. If two-way communication
is possible, the protocol can be followed by new stages, possibly affecting more qubit pairs. 

In the description of code based protocols a number of commuting joint measurements is performed by
Alice and Bob (corresponding to measuring the syndrome of a stabilizer code). The measurement
outcomes are compared in the same way as for the permutation based protocols. To end up with Bell
states a final operation is needed at both sides that corresponds to the decoding operation of the
stabilizer code. 

We show that both approaches are in correspondence in the following way. If the
measurement of the code based protocol is performed by first decoding, then doing qubit
measurements and then encoding again, the protocol becomes equivalent to a permutation based
protocol, as the encoding operation is canceled by the final decoding operation of the code based
protocol. The decoding operator of a stabilizer code applied at both sides yields a
permutation of tensor products of Bell states. Conversely, every permutation based protocol can be
interpreted as a code based protocol.

In this paper we will focus on two schemes presented in Refs.~\cite{DVD:03, M:03} that represent the two approaches.
In section~\ref{preliminaries}, we give an overview of some preliminary definitions and theorems. In section~\ref{schemes}, we describe the main results of the permutation based and the code based protocols, omitting the more detailed elaborations, as stated in Refs.~\cite{DVD:03, M:03}. Both of them are slightly moderated in order to clearly show the equivalence, which is done in section~\ref{equivalence}.

\section{Preliminaries}\label{preliminaries}
\subsection{Binary vector representation of products of Bell states}
Bell states can be represented by assigning two-bit vectors to the Bell states as follows
\begin{equation}
\begin{array}{ccccc}
\phiplus&=&\frac{1}{\sqrt{2}}(|00\rangle+|11\rangle)&=&|B_{00}\rangle\\
\psiplus&=&\frac{1}{\sqrt{2}}(|01\rangle+|10\rangle)&=&|B_{01}\rangle\\
\phimin&=&\frac{1}{\sqrt{2}}(|00\rangle-|11\rangle)&=&|B_{10}\rangle\\
\psimin&=&\frac{1}{\sqrt{2}}(|01\rangle-|10\rangle)&=&|B_{11}\rangle.
\end{array}\end{equation}
A tensor product of $n$ Bell states can then be described by a $2n$-bit vector, e.g. $|B_{010011}\rangle=|B_{00}\rangle|B_{11}\rangle|B_{01}\rangle=\phiplus\psimin\psiplus$. Note that the first half of the subscript refers to the phase of the Bell states (''$+$'' or ''$-$''), and the second to $|\Phi\rangle$ or $|\Psi\rangle$.

An interesting feature is the correspondence between Bell states and Pauli matrices
\begin{equation}\label{bellpauli}\begin{array}{rcl}
\phiplus&\rightarrow&\frac{1}{\sqrt{2}}\sigma_{00}=\frac{1}{\sqrt{2}}\sigma_0=\frac{1}{\sqrt{2}}\left[\begin{array}{cc}1&0\\0&1\end{array}\right]\\
\psiplus&\rightarrow&\frac{1}{\sqrt{2}}\sigma_{01}=\frac{1}{\sqrt{2}}\sigma_x=\frac{1}{\sqrt{2}}\left[\begin{array}{cc}0&1\\1&0\end{array}\right]\\
\phimin&\rightarrow&\frac{1}{\sqrt{2}}\sigma_{10}=\frac{1}{\sqrt{2}}\sigma_z=\frac{1}{\sqrt{2}}\left[\begin{array}{cc}1&0\\0&-1\end{array}\right]\\
\psimin&\rightarrow&\frac{1}{\sqrt{2}}\sigma_{11}=\frac{1}{\sqrt{2}}\sigma_y=\frac{1}{\sqrt{2}}\left[\begin{array}{cc}0&-i\\i&0\end{array}\right].
\end{array}\end{equation}
A tensor product of $n$ Bell states is then described by a kronecker product of Pauli matrices, e.g. $\phiplus\psimin\psiplus\rightarrow\frac{1}{\sqrt{8}}\sigma_0\otimes\sigma_y\otimes\sigma_x=\frac{1}{\sqrt{8}}\sigma_{00}\otimes\sigma_{11}\otimes\sigma_{01}$. In the same way we can use longer vector subscripts to denote such kronecker products, e.g. $\sigma_{010011}=\sigma_{00}\otimes\sigma_{11}\otimes\sigma_{01}$. All the $4.4^n$ tensor products of Pauli matrices multiplied with a global phase $\in\{\pm 1,\pm i\}$ form the \emph{Pauli group} ${\cal P}_n$.

The Pauli group and the group of binary vectors $\mathbb{Z}_2^{2n}$, are related to each other in the following way 
\begin{equation} \sigma_a\sigma_b\sim\sigma_{a+b}, \end{equation} where ''$\sim$'' denotes equality up to a global phase \cite{D:03}. Such a phase is irrelevant when these matrices represent pure state vectors. All addition of binary objects is done modulo 2.

Note that any two elements of the Pauli group either commute or anticommute. It can be easily verified that two Pauli matrices $\sigma_a$ and $\sigma_b$ commute if the \emph{symplectic inner product} $a^TPb$ is equal to zero, or 
\begin{equation}\sigma_a\sigma_b=(-1)^{a^TPb}\sigma_b\sigma_a, \quad\mathrm{where}\quad P=\left[\begin{array}{cc}0_n&I_n\\I_n&0_n\end{array}\right].\end{equation}

\subsection{Local permutations of products of Bell states}
In general a pure bipartite state $|\psi\rangle$ of $2n$ qubits can be represented by a $2^n\times 2^n$-matrix $\psitilde$ (e.g. as in (\ref{bellpauli}) for Bell states). Local unitary operations $|\psi\rangle\rightarrow(U_A\otimes U_B)|\psi\rangle$, in which Alice acts on her $n$ qubits (jointly) with an operation $U_A$ and Bob on his $n$ qubits with an operation $U_B$, are then represented by
\begin{equation}\psitilde\rightarrow U_A\psitilde U_B^T.\end{equation}
Interesting local unitary operations for the protocols under consideration are those that result in a permutation of the $4^n$ tensor products of $n$ Bell states (up to a global phase). In Ref.~\cite{DVD:03} it is proven that if local unitary operations result in a permutation of the products of Bell states, this permutation can be represented in the binary vector representation as an affine symplectic operation
\begin{equation}\label{permutatie}\begin{array}{ll} & \phi:\mathbb{Z}^{2n}_2\rightarrow\mathbb{Z}^{2n}_2:Ax+b\\
\mathrm{with} & A \in \mathbb{Z}^{2n\times2n}_2, b \in \mathbb{Z}^{2n}_2\\
\mathrm{and} & A^TPA=P\end{array}\end{equation} 
and that any such permutation $\phi$ can be realized by local unitary operations 
\begin{equation}\label{permutatie2} \sigma_x\rightarrow U_A\sigma_x U_B^T=\sigma_{Ax+b}.\end{equation} For an efficient way of doing this (by means of $O(n^2)$ one and two-qubit operations), we refer to Refs.~\cite{DVD:03, D:03}. A matrix satifsying (\ref{permutatie}) is called \emph{P-orthogonal} or \emph{symplectic}.

The examined protocols result in a new mixture of tensor products of Bell states
\begin{equation}\label{resultstate}\rho=\sum\limits_{x\in\mathbb{Z}^{2m}_2}p_x|B_x\rangle\langle B_x|.\end{equation}
The maximum value $p_a$ of the $p_x$ in (\ref{resultstate}) can be used as an entanglement measure. Note that one can always transform $p_0$ into this maximum value by applying at either Alice's or Bob's side the corresponding Pauli operator $\sigma_a$ that rotates $|B_a\rangle\langle B_a|$ into $|B_0\rangle\langle B_0|$. We will call $p_0=\langle B_0|\rho|B_0\rangle$ the \emph{fidelity} $F$.

\subsection{Stabilizer codes}
A stabilizer $S$ is a commutative subgroup of the Pauli group ${\cal P}_n$ which does not contain $-1$ or $\pm i$. The stabilizer code ${\cal H}_S\subseteq{\cal H}_{2^n}$ associated with $S$ is the \emph{joint eigenspace} of all elements of $S$, or 
\begin{equation} |\psi\rangle\in{\cal H}_S\Leftrightarrow M|\psi\rangle=|\psi\rangle,\quad\forall M\in S.\end{equation}
The group $S$ is called the stabilizer of the code, since it preserves all of the states in the code. If the stabilizer has $n-m$ independent generators $M_i$, the code space has dimension $2^m$ \cite{G:00}. From the definition of $S$ it follows that the eigenvalues of each $M_i$ are $+1$ and $-1$, with the same multiplicity.

Suppose an error $E\in{\cal P}_n$ has afflicted a state $|\psi\rangle\in{\cal H}_S$. If $E$ and $M_i$ commute, then 
\begin{equation} M_iE|\psi\rangle=EM_i|\psi\rangle=E|\psi\rangle,\end{equation} so the error preserves the value $+1$ of $M_i$. If $E$ and $M_i$ anticommute, then \begin{equation} M_iE|\psi\rangle=-EM_i|\psi\rangle=-E|\psi\rangle,\end{equation} so that the error flips the value of $M_i$, and the error can be detected by measuring $M_i$. Repeating this procedure for every generator $M_i$ of $S$, we may write 
\begin{equation}M_i E=(-1)^{s_{i}}EM_i.\end{equation} The $s_{i}$, $i=1,\ldots,n-m$ constitute a \emph{syndrome} for the error $E$ as $(-1)^{s_{i}}$ will be the result of measuring $M_i$ if the error $E$ has occurred. For recovery, a Pauli operator $R$ is applied that has the same commutation relations with the $M_i$ as $E$. Recovery is successful if $RE\in S$, in which case error + recovery has a trivial effect on states $\in{\cal H}_S$.

Measuring the syndrome $s$ comes down to projecting onto the joint eigenspace of the $M_i$ with eigenvalues $s_i$. Note that this eigenspace is the code space associated with the stabilizer with generators $(-1)^{s_i}M_i$. We will call this code space ${\cal H}_{S,s}$.

\section{Two schemes for creating entanglement distillation protocols}\label{schemes}
\subsection{Permutation based protocols}
A slightly generalized variant of the distillation protocols presented in Ref.~\cite{DVD:03} can be summarized as follows. 
\begin{enumerate}
\item Start from a mixture of $4^n$ tensor products of Bell states. Typically, this is the tensor product of $n$ identical independent Bell diagonal states.
\item Apply a permutation of these $4^n$ products of Bell states with local unitary transformations as described in the preceding section. 
\item Check whether the last $n-m$ qubit pairs are $|\Phi\rangle$ or $|\Psi\rangle$-states. This can be accomplished locally by measuring both qubits of each pair in the $|0\rangle$, $|1\rangle$ basis, and checking whether both measurements yield the same or the opposite result. 
\item\label{sigma_a} Perform $m$ single-qubit Pauli operations to Bob's remaining qubits as described further. This comes down to rotating the resulting state (\ref{resultstate}) where $p_a$ is maximal to a state where $p_0$ is maximal.
\item If the resulting state of the $m$ remaining pairs satisfies a certain criterion (e.g. the fidelity exceeds a certain proposed value), keep them, otherwise, discard them. The result is a new mixture of $4^m$ products of Bell states.
\end{enumerate}
Using the same techniques as in Ref.~\cite{DVD:03}, one obtains the following. If Alice and Bob apply the above protocol, starting from an initial state
\begin{equation}\label{initial}\sum\limits_{x\in\mathbb{Z}^{2n}_2}p_x|B_x\rangle\langle B_x|,\end{equation}
the resulting state of the $m$ remaining pairs after the measurement is
\begin{equation}\label{result} 2^{n-m}\sum\limits_{y\in\mathbb{Z}^{2m}_2}\left(\frac{\sum_{x\in{\cal S}+PA^TP\bar{y}}p_x}{\sum_{x\in{\cal S}^{\perp}+PA^TP\bar{0}}p_x}\right)|B_y\rangle\langle B_y|\end{equation} where ${\cal S}$ is the subspace spanned by the rows of $AP$ with indices $n+m+1,\ldots,2n$ and ${\cal S}^{\perp}$ is the subspace of all the binary vectors that have a symplectic inner product equal to zero with the elements of ${\cal S}$. Note that ${\cal S}\subset{\cal S}^{\perp}$. $\bar{y}\in\mathbb{Z}^{2n}_2$ is constructed from $y\in\mathbb{Z}^{2m}_2$ as follows 
\begin{equation}
\bar{y}=\underbrace{y_1 y_2\ldots y_m}_m\underbrace{00\ldots 0}_{n-m}\underbrace{y_{n+1} y_{n+2}\ldots y_{n+m}}_m\underbrace{t_1 t_2\ldots t_{n-m}}_{n-m},\end{equation} where $t$ is the outcome of the measurements of the last $n-m$ qubit pairs, i.e. zeros where the measurements are the same, ones where they are opposite.

In step~\ref{sigma_a}, Bob applies to his remaining $m$ qubits the $\sigma_a$ for which the coefficient of $|B_a\rangle\langle B_a|$ in (\ref{result}) is maximal, or 
\begin{equation}a=\arg\!\!\!\max\limits_{y\in\mathbb{Z}^{2m}_2}\left(\sum_{x\in{\cal S}+PA^TP\bar{y}}p_x\right).\end{equation} Note that a solution is not unique: any element in the coset $a+{\cal S}$ will do. The resulting fidelity is then 
\begin{equation}\label{fid} F=2^{n-m}\left(\frac{\sum_{x\in{\cal S}+PA^TP\bar{a}}p_x}{\sum_{x\in{\cal S}^{\perp}+PA^TP\bar{0}}p_x}\right).\end{equation} 

\subsection{Code based protocols}
The protocols described in Ref.~\cite{M:03} are derived from a stabilizer $S$ with generators $\sigma_{g_i}$ for $i=1,\ldots,n-m$. In Ref.~\cite{M:03}, the protocols are more generally defined for higher-dimensional systems (qudits). We only consider qubits here. The protocol goes as follows: 
\begin{enumerate}
\item Alice measures the observable $\sigma_{g_i}^{\ast}$ for each $i$, where ''$\ast$'' stands for elementwise complex conjugation. Let $(-1)^{a_i}$ be the results of the measurements: they are the eigenvalues of the stabilizer code space ${\cal H}_{S^{\ast},a}$ containing the state after measurement. 
\item Bob measures the observable $\sigma_{g_i}$ for each $i$. Let $(-1)^{b_i}$ be the eigenvalues of the stabilizer code space ${\cal H}_{S,b}$ containing the state after measurement. 
\item\label{step} Bob performs the error correcting process treating the string $s=(b_1+a_1,\ldots,b_{n-m}+a_{n-m})$ ($\mathrm{mod}\ 2$) as a syndrome as described below.
\item\label{inversecoding} Alice applies to her share of the qubit pairs the inverse of the encoding operator of ${\cal H}_{S^{\ast},a}$. Bob applies the inverse of the encoding operator of ${\cal H}_{S,b}$ to his share.
\item\label{endstep} They discard the last $n-m$ qubit pairs. 
\item If the fidelity of the resulting state (which depends on the difference of the measurement outcomes $(b_1+a_1,\ldots,b_{n-m}+a_{n-m})$) of the $m$ remaining pairs is large enough, they are kept, otherwise, discarded.
\end{enumerate}
In step~\ref{step}, Bob has to apply a recovery operator $R$ treating the string $s$ as a syndrome. $R$ is defined as follows. Let $C$ be the linear subspace of $\mathbb{Z}_2^{2n}$ generated by $g_1,\ldots,g_{n-m}$, and $C^{\perp}$ the orthogonal space of $C$ with respect to the symplectic inner product. Choose $v\in\mathbb{Z}_2^{2n}$ such that $\sigma_v$ has commutation relations with the $\sigma_{g_i}$ corresponding to $s$, or $v^TPg_i=s_i$. Now define $u$ as the vector having maximum 
\begin{equation} \sum\limits_{x\in C+u}p_x\end{equation} in the coset $C^{\perp}+v$. Maximum fidelity (after step~\ref{endstep}) is achieved if Bob applies $R=\sigma_u$ in step~\ref{step} and is equal to 
\begin{equation}\label{fid2} F=2^{n-m}\left(\frac{\sum_{x\in C+u}p_x}{\sum_{x\in C^{\perp}+v}p_x}\right).\end{equation}

\section{Equivalence of the two approaches}\label{equivalence}
We will first show that measuring the observables in the code based protocols can be carried out by local unitary operations that yield a permutation of the tensor products of the Bell states, followed by single qubit measurements and the inverse of the local operations. Second, we show more specifically that the protocols of Refs.~\cite{DVD:03} and \cite{M:03} have the same results, i.e. the formulas for the resulting fidelity in Refs.~\cite{DVD:03} and \cite{M:03} are equivalent.

\subsection{Observable measurements in the code based protocols}
Let us first consider Bob's actions. A way of measuring the observables $\sigma_{g_i}$ is by first applying $U^{\dag}$, which is the inverse of the unitary coding operator $U$, then measuring the last $n-m$ qubits in the $|0\rangle,|1\rangle$ basis, and then applying $U$. For every $b\in\mathbb{Z}^{n-m}_2$, $U$ transforms states
\begin{equation}|\underbrace{00\ldots 0}_{m}b_1b_2\ldots b_{n-m}\rangle,\ldots,|\underbrace{11\ldots 1}_{m}b_1b_2\ldots b_{n-m}\rangle\end{equation} to a basis of the stabilizer code space ${\cal H}_{S,b}$. $U$ satisfies
\begin{equation}\label{U}\sigma_{g_i}=U\sigma_{e_{m+i}}U^{\dag},\quad i=1\ldots n-m,\end{equation}
where $e_{m+i}\in\mathbb{Z}_2^{2n}$ has 1 in position $m+i$ and zeros elsewhere, or $\sigma_{e_{m+i}}=I^{\otimes(m+i-1)}\otimes\sigma_z\otimes I^{\otimes(n-m-i)}$. A way of efficiently implementing such coding operator $U$ is explained in Ref.~\cite{G:96}. The $U$ in (\ref{U}) and the inverse of the coding operator $U^{\dag}$ in step~\ref{inversecoding} of the protocol cancel each other, so Bob only applies $U^{\dag}$ and then measures his last $n-m$ qubits.

In the same way, we can see that, in order to measure the observables $\sigma_{g_i}^{\ast}$, Alice applies $U^T$ and then measures her last $n-m$ qubits. From $U_A=U^T$, $U_B=U^{\dag}$ and (\ref{U}) it follows that
\begin{equation}\sigma_{e_{m+i}}=U_A\sigma_{g_i}^{\ast}U_B^T.\end{equation}
If we interpret the $\sigma$ in this equation as representing tensor products of Bell states, we see with (\ref{permutatie}) and (\ref{permutatie2}) that the local unitary transformations $U_A$ and $U_B$ yield in the binary picture a permutation $\phi$, with 
\begin{equation} \phi(g_i)=e_{m+i},\forall i.\end{equation}
Note that $\sigma_{g_i}^{\ast}\sim\sigma_{g_i}$ and these matrices represent the same pure states as a global phase is irrelevant.

\subsection{Equivalence of the two schemes}
To point out that formulas~(\ref{fid}) and (\ref{fid2}) are equivalent, we have to show that the subspaces ${\cal S}$ and $C$ are the same and that the summations are done over the same cosets of ${\cal S}=C$ and ${\cal S}^{\perp}=C^{\perp}$. 

We define a symplectic matrix $B$ with $Be_{m+i}=g_i$, which means that colums $m+1$ through $n$ are $g_1,\ldots,g_{n-m}$. A matrix with these columns can always be completed to a symplectic matrix since $S$ is a \emph{commutative} group, and therefore we already have that $g_i^TPg_j=0,\forall i,j$. The inverse of this matrix is $B^{-1}=PB^TP$, which is also symplectic. $B^{-1}$ defines the permutation realized by $U_A$ and $U_B$. Note that fixing $B$ by adding columns to $B$ comes down to fixing bases of the stabilizer codes ${\cal H}_{S^{\ast},a}$ and ${\cal H}_{S,b}$. The last $n-m$ rows of $B^{-1}P=PB^T$ are $g_1,\ldots,g_{n-m}$, which means that ${\cal S}$ and $C$ in (\ref{fid}) and (\ref{fid2}) are the same.

We have $C^{\perp}+PB^{-T}P\bar{0}=C^{\perp}+v$ if $PB^{-T}P\bar{0}=B\bar{0}$ has symplectic inner products with $g_1,\ldots,g_{n-m}$ equal to $s_1,\ldots,s_{n-m}$. We know from the definition of $s$ and $t$ that $s=t$. We then have \begin{equation} B\bar{0}=\sum\limits_{i=1}^{n-m}s_i B_{n+m+i}, \end{equation} with $B_k$ the $k$-th column of $B$. We know from the symplecticity of $B$ that $B_{n+m+i}$ has symplectic inner product equal to zero with all the other columns of $B$ except $B_{m+i}=g_i$, from which follows that $(B\bar{0})^TPg_i=s_i$.

In the same way it can be verified that $\forall y\in\mathbb{Z}_2^{2m}:B\bar{y}\in C^{\perp}+B\bar{0}$, or $C+B\bar{a}\subset C^{\perp}+B\bar{0}$ so the cosets $C+B\bar{a}$ and $C+u$ are the same since they are both cosets of $C$ in $C^{\perp}$ having maximum total probability of their elements.

\section{Conclusion}
We have compared two approaches to the design of entanglement
distillation protocols. Although both approaches are based on
different concepts, they turn out to be to a large extent equivalent.

\begin{acknowledgments}
Dr. Bart De Moor is a full professor at the Katholieke Universiteit Leuven, Belgium. Research supported by: Research Council KUL: GOA-Mefisto~666, GOA AMBioRICS, several PhD/postdoc \& fellow grants; Flemish Government: FWO: PhD/postdoc grants, projects, G.0240.99 (multilinear algebra), G.0407.02 (support vector machines), G.0197.02 (power islands), G.0141.03 (Identification and cryptography), G.0491.03 (control for intensive care glycemia), G.0120.03 (QIT), G.0452.04 (new quantum algorithms), G.0499.04 (Robust SVM), research communities (ICCoS, ANMMM, MLDM); AWI: Bil. Int. Collaboration Hungary/ Poland; IWT: PhD Grants, GBOU (McKnow); Belgian Federal Science Policy Office: IUAP P5/22 (`Dynamical Systems and Control: Computation, Identification and Modelling', 2002-2006) ; PODO-II (CP/40: TMS and Sustainability); EU: FP5-Quprodis; ERNSI; Eureka 2063-IMPACT; Eureka 2419-FliTE; Contract Research/agreements: ISMC/IPCOS, Data4s, TML, Elia, LMS, Mastercard.
\end{acknowledgments}

\end{document}